\documentclass[Afour,sageh,times]{sagej}
\usepackage{moreverb,url}

\usepackage[colorlinks,linkcolor=black,bookmarksopen,bookmarksnumbered,urlcolor=black,citecolor=blue]{hyperref}
\usepackage{hyperref}
\usepackage{booktabs}
\usepackage{caption}
\usepackage{array}
\usepackage{caption}

\newcommand\BibTeX{{\rmfamily B\kern-.05em \textsc{i\kern-.025em b}\kern-.08em
T\kern-.1667em\lower.7ex\hbox{E}\kern-.125emX}}

\def\volumeyear{2025}

\begin{document}

\runninghead{Sen et al.}

\title{Ethical Frameworks for Conducting Social Challenge Studies}

\author{Protiva Sen\affilnum{1}, Laurent H\'ebert-Dufresne\affilnum{1}\affilnum{2}\affilnum{3}\affilnum{4}, Pablo Bose\affilnum{5}, Juniper Lovato \affilnum{1}\affilnum{2}\affilnum{3}}

\affiliation{
\affilnum{1}Department of Computer Science, University of Vermont, USA\\
\affilnum{2}Vermont Complex Systems Institute, USA\\
\affilnum{3}Complexity Science Hub, AT\\
\affilnum{4}Santa Fe Institute, USA\\
\affilnum{5} Department of Geography and Geosciences, University of Vermont, USA}

\corrauth{Juniper Lovato
University of Vermont,
Burlington, VT 05405, USA}

\email{jlovato@uvm.edu}

\begin{abstract} 
Computational social science research, particularly online studies, often involves exposing participants to the adverse phenomenon the researchers aim to study. Examples include presenting conspiracy theories in surveys, exposing systems to hackers, or deploying bots on social media. We refer to these as ``social challenge studies,'' by analogy with medical research, where challenge studies advance vaccine and drug testing but also raise ethical concerns about exposing healthy individuals to risk.
Medical challenge studies are guided by established ethical frameworks that regulate how participants are exposed to agents under controlled conditions. In contrast, social challenge studies typically occur with less control and fewer clearly defined ethical guidelines.
In this paper, we examine the ethical frameworks developed for medical challenge studies and consider how their principles might inform social research. Our aim is to initiate discussion on formalizing ethical standards for social challenge studies and encourage long-term evaluation of potential harms.
\end{abstract}


\maketitle

\section{Introduction}
\label{sec:intro}
In computational social science research, particularly in studies exploring online phenomena or technological systems, researchers often employ methodologies that involve exposing individuals or systems to potentially harmful conditions to gain insights into behaviors or vulnerabilities. 
Examples include participants in surveys of misinformation being exposed to conspiracy theories that they might not have known prior, or online social networks are deploying automated bots and exposing users to synthetic content, or exposing technological systems for attackers to test their resilience through adversarial interventions. Despite involving similar risks, the studies in these examples do not operate within a standardized ethical framework across disciplines. This absence creates a gap in addressing potential harms, assessing benefits, and ensuring accountability. 

A notable example of this tension arose when researchers submitted intentionally flawed patches to the Linux kernel without disclosure or consent to evaluate vulnerabilities of this open-source community, which led to calls for new ethics boards in computer science research \citep{dirksen2024don}. This study exemplifies the methodological and ethical challenges inherent in these types of studies. Although this approach can yield valuable insights, it also raises significant ethical questions about consent, harm, and societal benefit in computer science and computational social science.

A recent case study further highlights the importance of considering ethics in social challenge studies. In 2024, researchers from the University of Zurich conducted a large-scale experiment on Reddit's r/ChangeMyView forum \citep{404mediaResearchersSecretly}. They used AI-generated bots that pretended to be real people to persuade others in debates. These bots posted as pretending to be rape survivors, therapists, and crafted their responses based on what users have shared online. Over the course of a month, they posted approximately 1,700 AI-generated comments across 404 threads without anyone's knowledge or consent. After the study, the moderators and Reddit officials criticized the research, even bringing legal action. Similarly, in 2021, researchers at the University of Minnesota sent some seemingly harmless but flawed code patches to the Linux kernel project to show issues in how open-source code is reviewed \citep{holz2021ieee}. They did this without informing the maintainers or obtaining proper approval first, which raised questions among the open-source community and led to a push for changes to research ethics rules. Although the Reddit and Linux studies differed in their methods and communities, both revealed how research interventions conducted without consent or transparency can risk undermining trust and raise questions of safety and legitimacy. These cases highlight the need for more robust ethical frameworks that take into account the risks associated with studying real-world online and technical communities. While ethical questions related to exposing participants to psychological, social, or informational risks have long been examined within the social sciences, including sociology, psychology, and political science, where researchers have developed discipline-specific ethical frameworks and debated the limits of consent, deception, and acceptable harm \citep{hilbig2021bending, marzano2021covert, phillips2021ethics} these frameworks may benefit from complementary insights from other research traditions.

Challenge studies are a well-established method in medical research, providing critical insights into vaccine efficacy, drug development, and mechanisms of infectious disease transmission. In clinical research, challenge studies involve the deliberate exposure of healthy participants to pathogens with controlled risks, necessitating robust ethical frameworks to balance scientific progress with participant safety. Over decades, these frameworks have evolved to include rigorous standards for assessing risks, benefits, and long-term impacts on participants. This precedent has helped to ensure both the advancement of public health and the protection of the individuals involved in such research.

Our paper specifically examines a subset of computational social research which we call ``social challenge studies.''  While our paper draws heavily on medical challenge studies, we also recognize that the key ethical dilemmas, such as deception, exposing participants to psychological stress, and observing sensitive behaviors, have long been part of classic social science research \citep{milgram1963obedience, kelman1967deception, humphreys1970tearoom}. Recent work in psychology and research ethics has also shown that deception, debriefing, and consent remain active areas of debate, particularly in misinformation research and studies that involve manipulation without prior disclosure \citep{bohns2022toward, murphy2023conducting, verbeke2023informed}. Informed consent, even in these complicated and complex studies, is a crucial part of the ethical process. These studies show the long-standing challenge of balancing valuable insights with the risk of harm. In parallel, internet-research ethics communities have developed guidelines for studying behavior in networked environments, addressing platform consent, contextual integrity, and privacy expectations. They also highlight the practical limits of debriefing in large-scale or anonymous online studies \citep{aoir}. Understanding social-science debates through the lens of contemporary internet-research ethics provides important context for digital research and highlights the limits of relying solely on medical analogy frameworks. Accordingly, we examine the ethical implications of these studies in light of established frameworks developed for medical challenge studies as a complementary analytic perspective rather than a normative replacement for existing social science ethics frameworks. 

The use of medical challenge studies as a point of comparison is not intended to privilege biomedical ethics over social science frameworks, but rather to draw on a well-developed ethical vocabulary for reasoning about intentional exposure to risk, proportionality of harm, and spillover effects on third parties. The purpose of this work is not to create generalizable guidelines across all disciplines or methodologies, nor is it to review all existing ethical frameworks across fields, but to begin a discussion about what multidisciplinary ethical lessons can be learned from the medical field. Drawing on the ethical principles that govern medical challenge studies, we propose preliminary guidelines to address gaps in ethical oversight and risk assessment in social challenge studies. This effort aims to promote dialogue on the development of ethical practices that ensure both participant protection and responsible research. We start with the understanding that, even though the risks in these fields may look different (e.g., physical injury vs. psychological harm), the core ethical responsibility of balancing risk and benefit remains the same. In fact, several scholars argue that the traditional ethical distinctions between biomedical and social science research ethics are often overstated, since both areas share many core ethical concerns \citep{emmerich2016reframing, wassenaar2016learn}.

The structure of this paper is as follows. The section titled \textit{Medical Challenge Studies} reviews ethical frameworks for these studies. The section entitled \textit{Social Challenge Studies} examines case studies that illustrate the challenges of social research with controlled adverse conditions. The section entitled \textit{Lessons Learned: Applying Medical Recommendations to Social Challenge Studies} explores how principles from medical challenge studies can inform the development of ethical guidelines for social research. Finally, in our conclusion, we will consider the broader implications of these proposals and identify areas for future research.

\section{Medical Challenge Studies}
\label{sec:medicalCS}

\subsection{Definition}
A challenge study (also called a human challenge trial) can be defined as a clinical trial in which healthy participants are intentionally exposed to a test subject or pathogen in a controlled environment \citep{hope2004challenge,jamrozik2020human}. 
Human challenge studies frequently raise ethical concerns because they deliberately expose participants to risks that they would otherwise not face \citep{callaway2020should}. Although these studies offer significant scientific benefits, they require careful planning and rigorous ethical oversight to balance participant risks with potential public health benefits \citep{eyal2020human}. It is important to note that many of the standards we discuss here are not unique to medical challenge studies; rather, they originate from broader medical research ethics. Principles such as scientific justification, benefit and risk evaluation, informed consent, and independent ethical review were developed for human subjects research in general and were later applied to challenge studies. Acknowledging this broader foundation helps show that challenge studies build on an existing ethical framework rather than creating one from scratch.

We review some of the key components of ethical frameworks relevant to medical challenge studies by drawing on a multitude of sources. The standards discussed are not meant to be exclusive but to synthesize a large conceptual framework succinctly. We also outline these concepts in Table \ref{tab:table_example5}.

\subsection{Standards}

\begin{table*}
\small\sf\centering 
  \caption{Ethical Frameworks in Medical Challenge Study}
  \label{tab:table_example5}
   % \begin{tabular}{lll}
  \begin{tabular}{p{0.15\textwidth} p{0.52\textwidth} >{\raggedright\arraybackslash}p{0.25\textwidth}}
    \toprule
    \textbf{Ethical Framework} & \textbf{Guidelines} & \textbf{References}\\
    \midrule
     Scientific rationale & Study must address a significant gap in medical knowledge; It should have a well-defined hypothesis and research plan; Expected outcomes should provide benefits to science/public health. & \citep{bambery2016ethical, calina2020covid, franklin2001ethical, jamrozik2020covid,jamrozik2020human}\\
     \hline
     Public engagement & The required level of engagement should be proportional to the study's risks and societal relevance; Engage community to explain study's purpose and methodology; Address public concerns and incorporate feedback into study design; Share results openly with constituents and public; Balance transparency with participant confidentiality& \citep{jamrozik2020covid, jamrozik2020human, jamrozik2021human, world2020key,selgelid2018ethical} \\ 
     \hline
     Absence of alternative & Verify no other study methods can yield equivalent results; Document failed attempts to use alternative methods. &  \citep{jamrozik2020human, bambery2016ethical} \\
     \hline
     Appropriate organizer & Researchers must have expertise in challenge studies/med. research; Oversight committees should be in place to monitor the study. & \citep{calina2020covid,jamrozik2021human}\\
     \hline
     Independent ethical review (IRBs/RECs/REBs)& Ensure independent ethical review before the study starts participant recruitment;Verify that risks are minimized and proportionate; Review informed consent materials for clarity and completeness; Monitor ongoing safety reports and protocol deviations throughout the study. & \citep{gov1979belmontreport, edgar1995institutional, grady2015institutional, lapid2019institutional}\\
     \hline
     Informed consent & Clearly explain the study's purpose, procedures, risks, benefits; Ensure information is in understandable language/format; Allow time for participants to ask questions before deciding; Obtain consent with ethics board approval, where such review bodies are available; Offer a contact point for ongoing questions/concerns. & \citep{bambery2016ethical, calina2020covid, franklin2001ethical, hope2004challenge, jamrozik2020covid,jamrozik2020ethical,jamrozik2020human, katzer2023ethical, selgelid2018ethical,schaefer2020covid}  \\
     \hline
     Benefits and harms & Potential benefits should outweigh risks for participants/ society; Clearly outline potential benefits for participants and society; Assess both short-term and long-term harms; Conduct a thorough risk-benefit analysis prior to study initiation and, where applicable, independent ethical review. & \citep{calina2020covid,franklin2001ethical,jamrozik2020covid,selgelid2018ethical,katzer2023ethical,jamrozik2020ethical}\\
     \hline
     Risk minimization & Use lowest possible pathogen dose for efficacy; Provide immediate medical care in case of adverse events; Monitor participants closely during and after the study. &  \citep{bambery2016ethical,jamrozik2020covid,jamrozik2020ethical,gordon2017framework} \\
     \hline
     Third-party risks & Monitoring procedures to prevent pathogen spread beyond participants; Communicate third-party risks transparently during consent. &  \citep{jamrozik2020covid, jamrozik2020ethical, jamrozik2020human} \\
     \hline
     Selection of participants & Avoid discrimination on the basis of:  gender, race, socioeconomic status, or other irrelevant factors; Screen participants for physical/mental health to ensure suitability; Avoid selecting participants based on vulnerability/inability to refuse; Justify the need for any specific subpopulation in the study. & \citep{bambery2016ethical,jamrozik2020human,selgelid2018ethical,katzer2023ethical} \\
     \hline
     Payment of participants  & Ensure payments are proportional to time, effort, and inconvenience; Avoid offering excessive compensation that could influence decisions; Provide clear and transparent payment guidelines to participants. &  \citep{franklin2001ethical,bambery2016ethical,jamrozik2020human,katzer2023ethical,jamrozik2020ethical} \\ 
     \hline
     Right to withdraw & Participants should be able to leave the study at any time without penalty; Ensure participants are aware of their rights during the consent process; Provide support and follow-up care for those who withdraw. & \citep{franklin2001ethical,jamrozik2020human, jamrozik2021human, kenneally2012menlo,schaefer2010right} \\
  \bottomrule
\end{tabular}
\end{table*}

\subsubsection{Scientific Rationale.}
A standard of scientific justification requires sufficient evidence from researchers that the benefits of research, such as public health advancement or policy information, outweigh the associated risks \citep{calina2020covid, franklin2001ethical, bambery2016ethical}. This standard also guides the design of reliable and impactful studies that yield findings of value to both the scientific community and society at large \citep{world2020key}.

\subsubsection{Public Engagement.}
Public engagement is essential in medical science, particularly in challenge studies, as it fosters transparency, collaboration, and trust \citep{jamrozik2020covid}. Early public participation aligns the research process with societal realities and interests, ensuring that community needs are considered in the design and conduct of the study \citep{world2020key, selgelid2018ethical}. Transparent and open communication about the aim, approaches, and potential risks and benefits of the study promotes understanding and trust \citep{jamrozik2020human}. The impact of these activities would be further maximized by regularly updating community representatives with newly available information that reflects broader research initiatives and public health strategies \citep{world2020key,jamrozik2020covid}. While public engagement plays an important role in many challenge studies, it is not necessary for all of them. The degree of public engagement required for challenge studies should be proportional to the study’s associated risks and its public health relevance.

\subsubsection{Absence of Alternatives.} The use of medical challenge studies often depends on the lack of other plausible options to achieve the research objectives. 
However, this requires evidence that there is no reasonable alternative to achieve the objectives of the study, ensuring that the ethical and scientific merits of the study exceed the associated risks \citep{bambery2016ethical}.

\subsubsection{Appropriate Organizer.} For conducting medical challenge studies, institutions with appropriate organizational ability are required for participant safety and the validity of the study. They should have the necessary specialized infrastructure \citep{calina2020covid}, such as containment units or emergency care resources, to handle the risks associated with the exposure of participants to pathogens and for the handling of personally identifiable information. 

\subsubsection{Independent Ethical Review.} Independent ethical review represents an important mechanism for supporting ethical research involving human participants. This process is carried out by an impartial body, often called a Research Ethics Committee (REC), Institutional Review Board (IRB), or Research Ethics Board (REB) internationally, whose role is to assess whether a study meets accepted ethical standards. The primary purpose of this committee is to ensure the research design aligns with fundamental ethical principles, specifically: maximizing potential social value, minimizing harm to participants and communities, and ensuring the fairness and voluntariness of the informed consent process \citep{gov1979belmontreport, edgar1995institutional}.

Before a study can begin recruiting participants, the committee usually performs a thorough review to confirm that the risks are as low as possible and that any remaining risks are reasonable with respect to the potential benefits of the study \citep{grady2015institutional}. As part of this process, they also check informed-consent materials to ensure that they clearly explain the study, its possible risks, and the rights of participants in a way that anyone can understand \citep{lapid2019institutional}. While access to such independent ethical review varies across institutional, disciplinary, and geographic contexts, this form of oversight remains a valuable though not universal component of research governance, complementing researchers’ responsibility to adhere to established ethical frameworks and professional norms.

\subsubsection{Informed Consent.} One of the most important ethical principles in research is informed consent, which ensures that potential benefits must outweigh the risks of harm to both subjects and society
 \citep{schaefer2020covid, jamrozik2020ethical}. It is even more pertinent for challenge studies, in which participants are deliberately exposed to risks. Challenge studies require clear and detailed explanations of the purpose of a study, along with a description of the risks involved and a discussion of their potential long-term consequences \citep{franklin2001ethical,jamrozik2020ethical}. Obtaining valid informed consent involves providing not only thorough explanations of the research but also an understanding of the possible health risks and the potential broader societal benefits that might result from it. Participants must agree voluntarily or willingly to be part of the research, understand the possible consequences, and have enough time to make an informed decision without coercion or undue pressure \citep{hope2004challenge}. 

However, it can be argued that in certain cases where the public interest is at stake, it is ethically permissible to involve study subjects without obtaining informed consent. Such an approach is also viable if providing informed consent is not possible and if the participation of the subject does not impact their autonomy. In all cases, the ethical justification of such decisions must be carefully evaluated \citep{resnik2018ethics, gelinas2016and}. In certain situations, researchers may require a waiver of informed consent from an Institutional Review Board (IRB) before beginning their study. It should be noted that IRBs (or their equivalents) do not exist to provide ethical oversight in all research contexts; indeed, these are far more normalized in North America than in other parts of the globe. Disciplinary associations often also provide ethical guidance in many cases. 

When an IRB is available, a consent waiver can be obtained. This waiver is necessary to protect the interests of the participants and ensure that the research is conducted ethically. It is important for researchers to understand the importance of obtaining a waiver from an Institutional Review Board (IRB), as it is pivotal to protecting the rights of study participants. To obtain a waiver of informed consent researchers have to ensure that: (1) the research involves no more than minimal risk to the subjects; (2) the waiver or alteration will not adversely affect the rights and welfare of the subjects; (3) the research could not practicably be carried out without the waiver or alteration; and (4) whenever appropriate, the subjects will be provided with additional pertinent information after participation((Department of Health and Human Services 2009 at 45 CFR 46.116d) \citep{regulations2009protection}.

\subsubsection{Benefits and Harms.} In human challenge studies, the potential benefits must outweigh the risks of harm to both subjects and society. Participants may benefit from these interventions if they receive the most modern medical treatment and the option of possible early treatment. Society, on the other hand, may benefit from the rapid advancement of research, such as the development of treatments or interventions \citep{jamrozik2020covid,schaefer2020covid}. It is important to note that while some challenge studies may offer individual benefits (e.g., acquired immunity or access to high-quality care), others do not provide direct benefits to participants. The primary anticipated benefits of these studies are societal rather than individual.

However, it is also important to consider short- and long-term adverse effects. In medical research, short-term harms include adverse reactions to exposure to the pathogenic agent, whereas long-term consequences could be permanent functional disabilities or death resulting from an infection. 

In addition to physical harm, participants may also face burdens that affect their well-being, such as commitment to time, inconvenience, and psychological stress. Ethical frameworks for human subjects research emphasize that both harms and burdens should be incorporated into risk–benefit assessments \citep{bambery2016ethical, emanuel2000makes}. Therefore, it is imperative to perform a detailed risk-benefit analysis before any study is approved to minimize the risk of harm and to justify the expected benefits to participants and society \citep{bambery2016ethical,franklin2001ethical}. It is also essential to evaluate risk longitudinally and implement strategies to contact challenge study participants retrospectively to assess any harm they may have experienced.

\subsubsection{Risk Minimization.} Minimization of risk to participants means incorporating measures that reduce the risk of harm to participants while not compromising the quality of research. This can be achieved by appropriate participant selection, such as healthy subjects with particular and defined eligibility criteria, and by excluding vulnerable groups \citep{schaefer2020covid}. More specifically, strict safety measures, including continuous surveillance of participant health and rapid availability of medical care in the event of side effects, should be ensured \citep{hope2004challenge}. Research design also needs to be driven by reducing contact with pathogens and by using the least invasive techniques for research purposes. Informed consent procedures should allow participants to understand the possible risks and participate of their own free will \citep{jamrozik2020covid,jamrozik2020ethical}.

\subsubsection{Third-party Risk.} Third-party risks in challenge studies involve possible harm to people or communities who do not participate in the research. These risks can arise from unexpected outcomes, such as pathogens that escape controlled settings \citep{jamrozik2020covid}. These concerns are more pressing in endemic or low- and middle-income countries, where public health networks might not be able to handle and stop surprise outbreaks \citep{jamrozik2020human}. In endemic settings with high levels of background risk, risk assessment is complicated due to the fact that the marginal risk from participation in challenge studies is likely low compared to the background, yet the absolute risks for participants remain high. In low-resource settings, measures to mitigate additional risks might not always be available. To address these issues, scientists need to put in place strong safeguards. These include keeping participants isolated when they are infectious, following strict biosafety rules, and making sure the laboratories are well-contained. Ethical guidelines require a thorough check of the risk of harm to third parties during planning and approval, with clear plans to reduce risks built into the study design \citep{jamrozik2020ethical}.

\subsubsection{Selection of Participants.} The selection of participants for human challenge studies should emphasize fairness and minimize harm, particularly by focusing on young healthy adults (18-30 years old). These parameters, of course, can change depending on the study parameters and risk factors \citep{calina2020covid,world2020key}. 
These participants would have a low probability of background infection, minimizing any additional risks related to the study \citep{jamrozik2020ethical}. Participants who are at high risk for infection due to social injustice should be excluded from the study. It would be unethical and could take advantage of their vulnerabilities \citep{world2020key}. Vulnerable populations, such as children, pregnant individuals, and those with preexisting health conditions, are often excluded from studies to prevent exacerbating inequalities or causing additional harm to these groups \citep{jamrozik2020covid,bambery2016ethical}. Ensuring equity in participant selection is essential to prevent the exploitation of disadvantaged groups, especially in low- and middle-income countries where socioeconomic inequalities might undermine voluntariness of consent \citep{jamrozik2020human,selgelid2018ethical}.

%\textbf{Payment of participants}
\subsubsection{Payment of Participants.} When considering the payment of participants in human challenge studies, it is important to ensure that financial incentives do not unduly influence their decision to participate. Compensation for individuals for their time and the inconveniences they face during the study is ethical \citep{calina2020covid}. However, payment amounts should not be so high that they create pressure, particularly for people from vulnerable or socioeconomically disadvantaged backgrounds \citep{bambery2016ethical}. 
Compensation should be aligned with the risks and demands of the study, ensuring fairness and avoiding exploitation \citep{franklin2001ethical, katzer2023ethical}. 
Researchers must avoid using payments to exploit economically vulnerable groups and ensure that compensation does not encourage participants to conceal risks to receive payment \citep{selgelid2018ethical, jamrozik2020ethical}.

%\textbf{Right to withdraw.}
\subsubsection{Right to Withdraw.} In challenge studies, participants must have the freedom to withdraw from the study at any time, without any harmful consequences or loss of any benefit \citep{franklin2001ethical,jamrozik2020human, kenneally2012menlo}. They should be well informed that they can withdraw from the study whenever they want, even after giving their consent. Such decisions should not involve coercion or disadvantage. The right of withdrawal respects the autonomy of the participants and protects them from undue pressure to continue with the study. They should also be assured that withdrawal from the study will not affect their access to medical care, compensation, or eligibility for future studies \citep{schaefer2010right}. In addition, researchers should maintain open communication so that participants feel comfortable sharing their concerns or requesting to withdraw without worrying about negative consequences \citep{kenneally2012menlo,schaefer2010right}.

\section{Social Challenge Studies}
\label{sec:SCS}

We define social challenge studies as research that deliberately introduces structured risks (for example, risks of deception, privacy loss, or the potential for perturbing group dynamics or institutional processes) to individual participants or to communities or social groups, in order to generate scientific knowledge. Here, ``risk'' refers to the possibility of harm arising from research interventions, recognizing that such risks may or may not materialize as actual harms during the course of a study. This includes, but is not limited to, studies of misinformation and conspiracy theories, studies that use bots in online platforms, some studies of open-source software, and certain cybersecurity studies. 

In this section, we outline areas of research we believe fit the criteria of a social challenge study. These studies typically adhere to the ethical guidelines of their field and are often developed by professional organizations \citep{aasCodeEthics,aagStatementProfessional, europeansociologyEuropeanSociological}. At the same time, these research domains are inherently multidisciplinary and may raise ethical questions, particularly regarding intentional exposure to risk, that are not always addressed in a systematic manner within existing disciplinary codes. For this reason, social challenge studies could benefit from a more comprehensive ethical framework that complements established social science and internet research ethics guidelines by incorporating insights from medical challenge studies, without supplanting those existing frameworks.

\subsection{Misinformation and Conspiracy Theories}
\label{sec:misinfo}
%Laurent 

Research on misinformation focuses on factually incorrect (or sometimes simply unproven) information that is actively spread by people who believe it to be true. Online social media platforms have become central channels for the rapid diffusion of conspiracy theories, while simultaneously providing researchers with large-scale data to analyze their spread and evolution. Conspiracy theories can be broadly defined as unproven explanations of historical events or other aspects of reality based on the role of secretive small groups with harmful goals \citep{brian1999conspiracy, nera2023so}.

Because there is an important social signaling function related to sharing and broadcasting these beliefs \citep{bergamaschi2023signaling}, misinformation can spread widely on online platforms. Progressive exposure to a social group and its related concepts, ideas, and stories leads people to adopt beliefs in conspiracy theories \citep{Williams2022beliefs}. Susceptibility to misinformation has been shown to be positively correlated with a lack of interpersonal trust and employment security \citep{goertzel1994belief}, as well as a lack of trust in experts \citep{imhoff2018using, nera2024conspiracy}. However, susceptibility has also been shown to depend on personal demographics \citep{lovato2024diverse} and, in the case of conspiratorial thinking, it is sometimes even \textit{positively} correlated with historical knowledge \citep{adams2006perceptions, nelson2010role}. Therefore, given the broad range of potential misinformation, the structure of the problem is not simply a minority intrinsically prone to believing falsehoods, but is instead a complex system in which some beliefs can spread from one individual to another in surprising ways. 

\subsubsection{Misinformation challenge studies.}

Research on misinformation has developed in multiple directions in recent years, and much of the empirical work has involved exposing human subjects to misinformation in various forms. By exposure, here we mean that participants are presented with stories or claims that may or may not be true and that can influence their personal beliefs. First, research on social psychology has been able to explore how marginal online communities have facilitated the spread of conspiracy theories as traditional social ties and norms break down \citep{klein2018anomie}. Many of these studies, including those cited above, rely on traditional survey methods and expose participants to different belief statements. Importantly, because these studies often want to analyze the role of conspiratorial thinking, education, related beliefs, and other factors, they often intentionally target different demographics. However, these studies are rarely longitudinal, so little is known about the risk associated with this level of exposure.

Second, research in computer science and sociotechnical systems has explored ways to classify and moderate content, or even actively help inform users of online social networks. The detection, verification, and mitigation of online misinformation have all been the subject of numerous recent reviews, given the vast number of contemporary studies covering these topics \citep{rohera2022taxonomy, aimeur2023fake, alghamdi2024comprehensive}. These systems can take many different forms, but very often include human intervention or supervision in their algorithm. Humans can be included in the labeling of training data for the automated classification of information. They can also be included later in the algorithm, either as experts or as standard users, to flag information in real-time on online social networks \citep{micallef2020role}. In either of these tasks, individuals can be exposed to misinformation to which they might be susceptible, and once again, little is known about the associated risks.

Third, and indirectly related to misinformation, social media bots have been designed to mimic human behavior and help the study of online social networks \citep{boshmaf2013design}. These social bots are known to be able to amplify certain messages or narratives \citep{aimeur2023fake}. Researchers may deploy bots to simulate behaviors like retweeting or liking posts to study how they influence the spread of information by collecting how real accounts interact with the bots and their posts \citep{monsted2017evidence}. These accounts are not controlled by researchers and potentially contain personally identifiable information. More importantly, this method raises concerns about the ethical implications of research that might unintentionally contribute to the spread of false or misleading content. We revisit this topic in more detail further.

\subsubsection{Unintended harms and recommendations.}

Human subjects are involved in many key research areas on misinformation: understanding the susceptibility of individuals, understanding the spread of misinformation, and designing interventions to reduce its harmful impacts on online social media. 
The risk that a subject adopts certain beliefs because they were exposed through the study platform is likely low, given that the individual may not be part of the relevant social group \citep{bergamaschi2023signaling} and may not have been systematically exposed to relevant concepts \citep{Williams2022beliefs}, but it is also likely non-zero.

A recent paper titled ``Do Survey Questions Spread Conspiracy Beliefs?'' tackles this exact issue and measures the potential harms of misinformation challenge studies \citep{clifford2023survey}. The study is motivated by the fact that misinformation is known to spread through repeated exposure, yet research on misinformation routinely exposes participants to it. The authors then conduct a pre-registered experiment embedded in a panel survey. The methodology consists of 18 conspiracy questions, based on questions and format consistent with previous studies in the field, but importantly, allowed researchers to follow participants over weeks. The results show a significant 3\% increase in conspiracy beliefs following exposure to the survey. Interestingly, this effect is mainly observed if the conspiratorial statements are presented with a simple agree-disagree format. The effect mostly disappears if participants are offered an explicit choice between a conspiracy theory and an alternative explanation.

Results show that exposing participants to misinformation can affect them, but also that simple steps can be taken in survey or labeling tasks to reduce potential risks. In addition to concerns about misinformation, there is a more comprehensive methodological concern that surveys can often reinforce categories and social outcomes that shape the realities they seek to measure. Researchers have shown that surveys can reproduce descriptive constructs of race \citep{martinez2025facecraft}, sexuality \citep{westbrook2022dangerous}, and other identity categories \citep{law2009seeing}. This raises questions about whether the surveys are truly fair and accurate. Even in regular online surveys, it is still important to consider whether they are accurate and whether they may cause any unexpected consequences \citep{evans2018value}. These studies show that exposure effects are not just conspiracy theories. They also have to do with how survey questions can create, strengthen, and normalize sensitive categories. This can have long-term implications for the survey participants and for society as a whole.

Accordingly, we can follow several recommendations. First, as we just saw, recent results show that presenting factual information as an alternative to misinformation can help reduce the effect of exposure \citep{clifford2023survey}. Second, other research methods could be considered, such as ethnographic or observational studies of online communities to find individuals involved in conspiratorial thinking rather than exposing random participants to misinformation \citep{polleri2022towards}. This can also serve to minimize risk. Third, in both surveys and labeling tasks, participants could be screened to ensure suitability and avoid selecting participants with known susceptibility, such as a lack of job security or other risk factors \citep{goertzel1994belief}. Fourth, longitudinal studies should be considered in order to assess long-term risks and enable meta-studies on the potential impact of misinformation research \citep{clifford2023survey}. Finally, researchers should be aware of the risks of making social categories seem more real than they are when designing surveys. They need to carefully think about the categories they use, avoid simple yes/no questions or formats that reduce complexity, and interpret the survey results with a broader understanding of social identity based on qualitative and historical contexts.

\subsection{Online Social Media Research and Bots}
\label{sec:bots}

Research on social media platforms involves studies of user-generated content and user behavior. Social challenge studies in this area may include the generation of new content generated by researchers to monitor user reactions (for example, to gauge the perceived credibility of posts \citep{li2023assessing}). Bots, therefore, present a particularly compelling opportunity due to their scalability and ability to simulate human-like behavior on a large scale. 
This unique capability makes bots valuable tools for studying phenomena such as information spread, but it also raises complex ethical considerations.

\subsubsection{Bot challenge studies.} Bots are commonly used in research to study how information spreads through online social networks \citep{monsted2017evidence}. They can simulate behaviors like retweeting or commenting. But in general, bots can also greatly affect the platform that hosts them.  Bots can amplify divisive messages, causing them to spread more quickly and broadly \citep{ferrara2016rise}. Bots can also contribute to the spread of misinformation by making false stories appear more credible through extensive sharing, creating a frequency illusion \citep{shao2017spread}. Bots are also valuable for examining user behavior. Bots can influence the polarization of online discussions, emphasizing their role in the formation of digital echo chambers \citep{grimme2017social}. Furthermore, since bots act as pseudohumans, they intrinsically influence perceptions and trust dynamics in online interactions \citep{cresci2017paradigm}.

\subsubsection{Ethical issues in research involving bots.} Informed consent is a crucial ethical consideration in online social media research involving bots for two reasons. First, the methodology often relies on the fact that users will be unaware that they are interacting with bots \citep{krafft2017bots}. Second, bot-generated content is usually public, which means that researchers cannot control who interacts with it \citep{krafft2017bots}. 
If individuals do not know they are engaging with a bot, they cannot truly consent to the interaction, and their autonomy is being compromised. This raises concerns as respecting autonomy and informed decisions is a fundamental ethical principle in research. \citep{brock2008philosophical}.

Moreover, it is also a significant challenge to determine who is ethically and legally responsible for the actions of intelligent technologies when they cause harm \citep{coghlan2023chat}. In the case of online social media research that integrates social challenge studies, such as bot research, does liability lie with the researchers who created the bot, the platform that hosts it, or both? This challenge increases when bots are designed to function autonomously, making decisions in real-time based on user input or adapting to changing conditions, which may also lead to issues of interaction bias. 

Defined accountability frameworks are crucial for establishing responsibility for bot actions and addressing potential harm. Ethical oversight helps researchers identify and mitigate risks before deployment, while legal frameworks ensure accountability for any resulting harm. However, the lack of clear and consistent rules for using bots on social media creates significant ethical and legal challenges for researchers. Different social media platforms have varying policies: for example, X, which was formally known as Twitter \citep{twitter_2023}, permits some forms of automation, while Facebook \citep{facebook_2019} and YikYak \citep{yikyakinc_2025} strictly forbid it. These inconsistencies make it more challenging for researchers to assess the ethical implications of using bots in their studies. Violating Terms of Service raises ethical concerns and potential legal liabilities. According to the Computer Fraud and Abuse Act (CFAA) in the United States, accessing a platform in violation of its terms can be considered illegal \citep{vaccaro2015agree}. 

To address this issue, researchers should carefully consider whether they are adhering to platform policies and Terms of Service regarding consent. Researchers also need to ensure that ethical oversight boards, such as Research Ethics Committees (RECs), Institutional Review Boards (IRBs), or Research Ethics Boards (REBs), understand and approve this type of research.

\subsection{Cybersecurity and Honeypots}
\label{sec:honeypots}

\subsubsection{Cybersecurity challenge studies.}
Cybersecurity research focuses on protecting digital environments by studying vulnerabilities and malicious behaviors. Not unlike the open-source software example given in the Introduction, social challenge studies in cybersecurity often involve controlled exposure to threats to understand and mitigate risks. One common method is the use of honeypots: deliberately exposed systems or resources designed to lure attackers so that their behaviors can be observed and analyzed. Unlike traditional cybersecurity measures designed to deter threats, honeypots encourage interactions to gather information on attacker strategies and techniques~\citep{spitzner2003honeypots, provos2004virtual}. These resources can include physical systems, virtual environments, or abstract entities such as files and online accounts~\citep{onaolapo2016gmail, lazarov2016honey}. 

Honeypots do not create new malicious actors or necessarily increase attacker activity. Rather, they make existing malicious behavior more visible by presenting a system with which attackers are likely to interact. The intentional design and deployment of the honeypot to lure, isolate, and attract malicious actors is why we characterize this line of research as a social challenge study. By exposing these systems to potential threats (attackers), we can better understand attackers’ behavior and strategies in a controlled setting, but we also risk compromising the security of the environment. This is a unique case where human participants are the challenge, and the study environment itself is being challenged. It is therefore necessary to ensure that honeypot systems are designed, built, and deployed to minimize harm; e.g., by isolating potentially harmful honeypot environments.

\subsubsection{Recommendations for honeypots.}
Using honeypots presents significant ethical and technical challenges, particularly in terms of informed consent, deception, and risk management. In this line of research, the researcher must keep all recorded activity traces safe and avoid deanonymizing visitors to honey assets. Furthermore, if experiments involve running live malware samples (as seen in~\citep{onaolapo2016gmail}), adequate care must be taken to ensure that the malware samples do not harm any internal or external parties~\citep{rossow12:practices}. It is also important to protect the researcher responsible for honey assets. Honey assets and honeypot systems must be designed to minimize the potential harm that the researcher may face, e.g., if their identity were to become known during experiments. Hence, it is important to take advantage of Virtual Private Networks (VPNs) and proxies and to incorporate them into the honeypot infrastructure when necessary. Moreover, there can sometimes be conflicts between the requirements of minimizing harm (e.g., via sandboxing honey assets) and ensuring that those honey assets appear realistic. In line with the Menlo Report~\citep{kenneally2012menlo}, the right balance between these requirements must be found by the researcher.

Honeypot studies frequently conflict with the principles of informed consent, as attackers cannot be notified without affecting the research. Consent is typically waived in accordance with ethical guidelines such as the Menlo Report, which emphasizes minimizing harm and maximizing social benefit~\citep{kenneally2012menlo}. Honeypot experiments are in necessary conflict with the informed consent norms that are enforced by several regulations and professional codes \citep{emanuel2000makes, regulations2009protection}. They depart from conventional informed-consent norms because attackers cannot be notified or debriefed without undermining the study. However, waiver of consent can be considered justifiable when the risks are minimal, rights are not significantly violated, and the research addresses substantial public interest~\citep{gelinas2016and}. For example, studies that expose malicious actors to controlled honeypot environments seek to protect broader systems without unduly affecting the attackers themselves~\citep{regulations2009protection}.

In addition, honeypot studies raise ethical questions that challenge even the most adaptable guidelines. What is the researchers' duty to attackers? If attackers use fake items, do standard ethical rules still apply? These scenarios intentionally cause hostile actions with no opportunity to explain them later, creating an uneven ethical situation. Unlike medical research, where people agree to controlled harm for knowledge and safety, cybersecurity studies create ethical traps to study participants' actions without their approval, which goes against their best interests. Because of this, researchers must carefully measure possible harm to systems, themselves, and even the attacker.

Another challenge comes from research ethics committees (such as IRBs/REBs). Many research ethics committees are not equipped to evaluate honeypot or malware studies.  As a result, ethical review can become overly cautious or insufficiently critical. This gap highlights an opportunity for cooperation and knowledge sharing between cybersecurity experts and ethics reviewers, supported by guidelines like the Menlo Report, to better assess digital risks.

Ethical honeypot research faces several challenges, including insufficient education on computer ethics and technical risks. The lack of training and qualified instructors to teach ethics in computer security contributes to knowledge gaps among cybersecurity researchers, leaving many unaware of appropriate ethical frameworks \citep{stavrakakis2022teaching}. Ethical oversight boards are essential for reviewing research that involves human subjects. However, their limited understanding of advanced technological methods may make these studies difficult to review, which highlights the need for improved collaboration between researchers and review boards. This collaboration is crucial for ensuring compliance with ethical protocols. Technical risks also present significant concerns; for example, cybercriminals may exploit honeypot servers to access sensitive data, or participants may unintentionally expose personal information by linking their accounts to research systems \citep{onaolapo2019honeypot}. 

Considering the aforementioned issues, we propose that honeypot studies actually align with our ethical model. These studies often involve deception, consent waivers, potential harm to individuals or communities, and complexities in ethical oversight.  Similar to social bot experiments, honeypot studies raise questions about maintaining participant anonymity, what they are exposed to, how their data is used, and who is responsible.  This means that they should be included in an ethics framework that draws on lessons from medical challenge trials.

\section{Adaptability of Medical Ethics in Social Challenge Studies.}
When translating ethical lessons from medical studies to social studies, researchers must consider the conceptual similarities and differences in how studies are designed, the risks involved, and the necessary oversight. Ethical research with human participants, whether conducted in clinical settings or online, must balance the benefits of scientific discovery with the risks of potential harm to individuals and groups.  In this section, we draw clear comparisons between medical challenge studies and social challenge studies, pointing out how the ethical issues of medical research apply to social research and how to adjust for digital environments. Both types of studies involve exposing participants to potential risks: medical risks from pathogens in health studies and risks from bots, misinformation, or manipulated online experiences in online studies. Although the fields are different, the core ethical issues, such as exposure control, informed consent, and harm minimization, are closely related. These principles help researchers put ethical protections in place, even when they are not in a formal medical trial setting.

To support this framework, Table~\ref{tab:ethical_differences} presents a comparison of medical and social challenge studies, highlighting the particular ethical considerations required for online research. Understanding these differences makes it easier to see where adjustments are needed and where we can learn from similar situations.

\begin{table*}[t]
\small\sf\centering
  \caption{Key Differences Between Medical and Online Social Challenge Studies}
  \label{tab:ethical_differences}
  \begin{tabular}{lll}
    \toprule
    \textbf{Feature} & \textbf{Medical Challenge Studies} & \textbf{Social Challenge Studies}\\
    \midrule
    Research environment & Controlled lab/clinic & Uncontrolled, dynamic digital platform\\
    \hline
    Participant visibility & Identified, screenable individuals & Anonymous, dispersed users\\
    \hline
    Exposure Mechanism & Biological pathogen & Bots, misinformation, adversarial content\\
    \hline
    Risk Type & Physical (health-related) &	Cognitive, emotional, reputational, representational, social bias\\
    \hline
    Consent Process	& Formal, individual informed consent &	Often waived; may use terms of service or group consent\\
    \hline
    Oversight Infrastructure &	Institutional Review Boards (IRBs)	& ethics boards lack digital expertise\\
    \hline
    Containment Measures &	Physical isolation, medical intervention &	Rate limiting, platform moderation, sandboxing\\
    \bottomrule
\end{tabular}
\end{table*}

\subsection{Controlled Exposure.} 
In medical challenge studies, people are intentionally exposed to a virus or bacteria under controlled settings. This usually means that they receive a specific dose in a sanitized environment and that medical professionals are keeping a close eye on everything. The idea here is that this exposure is done for a clear scientific reason, with careful administration, and the dose amount is kept to what is necessary. However, social challenge studies also put participants in situations where they encounter harmful content, such as online bots, fake news, hate speech, or misleading information. Interestingly, some researchers use epidemiological methods to explain how misinformation spreads through people and communities, borrowing terminology from the study of viruses \citep{govindankutty2024epidemic}.

Like medical challenge studies, where the amount of pathogens given to participants is carefully managed, social studies must also carefully quantify the amount of misinformation or online research content participants encounter. This involves determining how often, how intensely, and for how long participants are exposed to harmful messages. For instance, if someone sees a mild false statement just once, it might not be a major ethical issue. But if they are exposed to repeated, emotionally charged lies, that would be unethical. This highlights the need for a more structured and systematic approach, similar to the one used in medical studies for dose-response relationships.  

These similarities do not imply equivalence; exposure in online social systems is less physiologically risky but often more diffuse and less predictable. The ethical relevance of the medical analogy lies in its articulation of how intentional exposure imposes specific types of ethical considerations on researchers, regardless of domain.

\subsection{Informed Consent.}
In medicine, informed consent is typically an individual act: participants are informed of the risks, given the opportunity to ask questions, and allowed to withdraw without penalty. Social research, however, often complicates this model. Data may be drawn from online platforms or group interactions where individuals are not directly approached, identities may be obscured, and researchers sometimes rely on terms of service as a proxy for consent. Yet, as both legal scholars and review boards point out, simply using a platform does not constitute meaningful agreement to participate in research \citep{straub2025participatory}. Controversial cases, such as the Reddit manipulation study, underscore how the absence of explicit consent, especially when deception is involved, raises serious ethical concerns \citep{404mediaResearchersSecretly}. It is important to note that even though some content or online ecosystems are publicly accessible online, ethical obligations differ when researchers enter semi-private or membership-restricted spaces. Community-specific norms, privacy expectations, and barriers to entry shape the forms of consent or moderator approval that are appropriate. Our framework treats online data as context-dependent rather than uniformly public. Users' contextual expectations of privacy, the norms of a particular community, and the presence of barriers to entry (e.g., application-only forums, moderated spaces, closed groups) all affect whether additional consent, community consultation, or platform-level approval maybe appropriate.

For networked or group-based data, the problem is sharper still. Consent is rarely confined to a single person: identifiable nodes can reveal information about others in the network, including secondary participants who never agreed to participate. This poses a fundamental tension between the individualistic model of consent and the relational nature of network data \citep{lovato2022limits, neal2024recommendations}.

There is no perfect solution, but several practices from the social sciences can guide ethical approaches. In closed populations, researchers can seek consent not only for individuals to share their own information but also permission for others to share information about them. In some contexts, consent may be gathered from a representative, such as a household head or community moderator, while still allowing space for individuals to opt out. When direct consent is impractical, researchers can provide clear public notice, establish straightforward opt-out mechanisms, and minimize risks to those indirectly implicated.

Across all cases, researchers must act as stewards of the people represented in their datasets. This means limiting deception, protecting privacy, and actively reducing the potential harms that fall on secondary participants. Group consent and representative consent should not be seen as substitutes for ethical responsibility but as part of a layered model of transparency, risk minimization, and respect for the relational character of social data.

\subsection{Risk Minimization and Mitigation.}
In medical challenge studies, having solid plans in place to keep participants safe is essential. Researchers take a close look at who they want to include, often leaving out anyone who has existing health problems that could make things riskier. After participants join the study, they are closely monitored in a controlled setting where anything can go wrong. If complications appear, they are ready with emergency treatments like antiviral drugs or oxygen to help right away.

Similarly, social challenge studies should have the necessary protections in place for online settings. The risks in these studies often revolve around how people think and feel, and what they might lose in terms of reputation, as well as their physical safety, as implications from online challenges can be taken and spread into offline contexts. Unlike medical challenge studies that address physical risk, social challenge studies must carefully consider broader social harms. This includes the possible creation or spread of stereotypes and biases. These can include reinforcing stereotypes, bias, or prejudice against certain groups \citep{cecchini2019reinforcing}. When researchers study sensitive issues like racism or sexism in real online environments, their work can unintentionally make these problems worse by normalizing biased behavior or spreading harmful ideas \citep{cecchini2019reinforcing, glaser2005prejudice}. For example, showing participants biased news could strengthen stereotypes, or the study’s results could be misunderstood in ways that unfairly stigmatize a group \citep{arendt2023media}. Therefore, ethical guidelines for social challenge studies should encourage researchers to think carefully about how their methods or findings might contribute to collective harm. In addition, to reduce bias and potential harm, researchers should include some helpful strategies, such as:
\begin{enumerate}
    \item Fact-checking tools that let users know when content might not be reliable.
    \item Alerts that inform users they are part of a study or interacting with something that is not real.
    \item Resources for psychological support, like options to opt out if users feel upset.
\end{enumerate}

However, not every study allows participants to access these protections immediately. Studies that rely on complete deception keep participants anonymous, in which participants do not even know they are being tested, and raise specific ethical issues. In these situations, participants cannot take protective measures, such as avoiding certain content or asking questions, leaving them exposed until the study is over or the truth emerges. Things get even more complicated in big online spaces like social media, where participants are often anonymous and unaware of their participation. This makes it difficult for researchers to follow up after the study to explain what happened, clear up any misunderstandings, or let people know they can pull their data. This violates ethical expectations established in both biomedical and behavioral research, where post-study debriefing is essential when deception is involved \citep{kenneally2012menlo}. When it is not possible to provide safeguards during the study, researchers should consider the following duty:

\begin{enumerate}
    \item Justify using deception and anonymity to an oversight board.
    \item Show that there is no less harmful way to conduct the study.
    \item Keep the exposure as limited in scope, frequency, and emotional weight as possible.
    \item When possible, prepare for a thorough debriefing after the study and let participants withdraw their data if they choose to, following ethical guidelines for digital research.
    \item Get approval from the community if you cannot get individual consent, for example, getting permission from the platform, platform admins, or moderators.
    \item Be open about how the study was done and its limits so that people can trust the findings.
\end{enumerate}

In summary, reducing risk involves the researcher being aware and responsible during their study design. This is especially true when participants cannot give their consent, opt out, or be informed later. In studies that involve deception or keep identities anonymous, researchers must make sure they minimize risks and harm, and they should follow up with proper accountability after the study.

\section{Lessons Learned: Applying Medical Recommendations to Social Challenge Studies}
\label{sec:ethicalframe}

\subsection{Objective}
Building on the earlier discussion about how medical ethics can inform social challenge studies, this section turns those ideas into practical guidance. Here, we lay out clear standards for designing, reviewing, and running these studies, drawing on both medical challenge research and established practices in social and computational work.
Through this exploration, we aim to contribute to the ongoing dialogue on creating multidisciplinary ethical frameworks that balance scientific innovation with accountability and the well-being of participants. 

Although medical challenge studies provide a natural reference point, we do not assume that standards developed for clinical research can be transferred wholesale to online social science research. Instead, the analogy rests on a narrower and more general normative premise: any form of research that deliberately introduces controlled adverse conditions in order to produce knowledge must satisfy a set of obligations concerning risk–benefit justification, harm minimization, informed consent or its justified waiver, and independent oversight. These ethical considerations reflect broad traditions in research ethics that span many fields. Medical challenge studies, however, operationalize these principles in particularly explicit ways by managing risk through deliberate exposure. Social challenge studies share this structural feature, even though the sources of harm, types of participants, mechanisms of exposure, and relevant communities differ. The medical challenge study framework, therefore, functions as a starting point for reasoning about how to structure protections in online social contexts, while requiring careful reinterpretation to account for digital settings, informational harms, autonomy constraints, and community-level effects.

We have summarized the applicable ethical frameworks in Table \ref{tab:ethical_frameSCC} and also created an additional decision tree (Figure \ref{fig:decisiontree_ethicalFW}) in the \textit{Appendix} to help researchers support the practical application of these frameworks. This decision tree guides researchers through key questions including whether their study involves intentional exposure to risk, whether deception is used, whether informed or group consent is feasible, and whether participants are vulnerable or likely to experience long-term risks. This design helps researchers identify when additional safeguards, such as longitudinal follow-up, post-exposure debriefing, or platform-level review, are ethically required. This decision tree translates broad ethical principles into a practical framework for diverse audiences.

\subsection{Standards Revisited}

\subsubsection{Scientific rationale.}

Drawing on both medical and social research ethics, social challenge studies require a robust scientific rationale to justify the importance of the study both for the research team and for any external ethical review, as well as its associated risks and benefits, for both participants and the general public. Such studies usually examine issues of system security, privacy, misinformation, and online manipulation, with the broader objective of promoting the public good through enhanced cybersecurity and improved user experience.
Researchers need to be transparent about their research goals and outcomes, enabling accountability among broader societal communities and ethical oversight boards. The scientific knowledge acquired resulting from social challenge studies must be both scientifically rigorous and impactful, while also ethically defensible, ensuring that progress respects the responsibilities to individuals and society. 

\subsubsection{Public engagement.}

Public engagement through community consultation is an important ethical consideration in social challenge studies. Although it should not replace individual informed consent, community consultation fosters trust and supports the ethical conduct of research by allowing investigators to engage with local constituents and subject matter experts and explain study objectives, gather feedback, and address concerns \citep{richardson2006role}. Community consultation highlights the importance of transparency and accountability, demonstrating a commitment to justice and beneficence. It can also lead to better research questions, help mitigate bias, and provide a clearer understanding of the research context, ultimately yielding more effective research results \citep{marshall2001ethical}. By actively involving representatives from the groups most likely to be affected by the study, researchers can more effectively identify and address potential risks or ethical issues that they (and even ethics review boards) might overlook. Ultimately, including diverse perspectives during the study planning phase can lead to a more equitable distribution of risks and benefits.

Defining what makes a ``community'' can be challenging because there is often no straightforward method to identify legitimate community members or guarantee meaningful participation. However, effective communication and collaboration can help address these issues. Researchers should provide background education on relevant health, social, or technological topics to empower community members to engage meaningfully in discussions about study design and the risks and benefits involved \citep{hintz2020best}. Collaborating with local organizations or community leaders can enhance engagement, fostering mutual respect and shared decision-making. These partnerships also acknowledge the cultural sensitivities of the groups involved and help 
to build long-term trust in the research enterprise \citep{hintz2020best, casari2023beyond}. Such collaborative approaches not only strengthen trust but also enable more informed and accurate research by uncovering otherwise hidden or neglected forms of 'dark data,' thereby broadening the evidence base for ethical and effective study design \citep{hand2020dark}.

Beyond the consultative phase, there is growing recognition that participants have the right to learn about study results, both as a form of respect for their autonomy and as an acknowledgment of their contributions \citep{fernandez2003informing, casari2023beyond}. Researchers should aim to communicate these results in clear, accessible language whenever possible, as long as it does not create undue risks to participants or investigators \citep{hintz2020best, carroll2023care}. In certain contexts, sharing of findings can be harmful or stressful if the information is unreliable, misrepresents the community, or exposes individuals to risks such as trauma, discrimination, or financial harm \citep{fernandez2003informing}. Furthermore, researchers studying sensitive or high-risk subjects (e.g., hate speech or cybercrime) must consider the potential for personal harm if they are identified and targeted by adversarial groups. Thus, decisions about how and when to disclose study outcomes should strike a balance between the ethical imperative of transparency and the duty to prevent harm to participants, communities, and investigators.

\subsubsection{Absence of alternatives.}

Social challenge studies should only be conducted when no reasonable alternatives exist to achieve similar research objectives. Before launching a full-scale study, researchers can use simulations, models, existing observational data, archival corpora, or controlled experiments to evaluate the potential effectiveness of the proposed intervention and to predict possible external effects. By conducting experiments in a controlled environment (either through computational modeling, sandboxes, or pilot studies), researchers can more accurately predict outcomes, enhance methods, and uncover unexpected risks before engaging real-world communities. 

Industry partnerships can foster an environment that minimizes risks for participants and third parties. Collaborations with online platforms, for example, enable researchers to conduct experiments in ``sandboxed'' accounts or restricted spaces that isolate the investigation from broader user populations \citep{leckenby2021sandbox, onaolapo2021socialheisting}. This approach allows for the testing of novel or potentially disruptive interventions without exposing full-scale user communities to potential harm. 

Mathematical, computational, and agent-based models can incorporate established theories and empirical parameters to explore how an intervention might unfold, offering insights into potential dynamics before any real-world implementation. Integrating emerging technologies, such as artificial intelligence simulations, provides an additional layer of potential risk mitigation. In certain cases (particularly those with potential adverse mental health implications), AI-driven modeling can help researchers anticipate and address potential harms, or perform potentially harmful experiments on AI subjects rather than human subjects \citep{park2024generative}. Using generative AI in research, on the other hand, introduces additional ethical issues, including consent, attribution, and bias, that should be carefully considered \citep{schlagwein2023chatgpt,davison2024ethics}.

In medical challenge studies, the ``absence of alternatives'' principle typically requires showing that no other method (e.g., animal models, in vitro studies, or simulations) can achieve the same results. However, for studies of social challenges, the assessment is more complex. Because researchers often have access to simulations, observational data, or archival corpora, it is rare that no alternative exists. Instead, the relevant ethical question is whether alternatives can provide knowledge that is equivalent in validity, scale, and ecological realism to what a live-challenge study would reveal. For example, while survey experiments can measure whether exposure to misinformation shifts individual beliefs, they cannot capture how repeated exposures through social networks interact with platform algorithms to drive large-scale diffusion. However, researchers should document how they evaluated alternatives, simulations, observational studies, or smaller pilot interventions, and explain why these approaches were inadequate to meet the objectives of their study.

\subsubsection{Appropriate organizer.}

For conducting social challenge studies, institutions should have the necessary specialized infrastructure, such as ethical research boards. In addition, individual researchers should possess expertise and training in conducting large-scale, hypothesis-driven experiments, have an understanding of what challenge studies entail, and receive training on the ethics of working with human subjects. 

\begin{table*}[ht!]
\small\sf\centering
  \caption{Ethical Considerations for Social Challenge Study}
  \label{tab:ethical_frameSCC}
  \begin{tabular}{ll}
    \toprule
    \textbf{Ethical Framework} & \textbf{Considerations for Social Challenge Studies} \\
    \midrule
     Scientific rationale &
     \shortstack[l]{Address emerging social concerns: such as misinformation, online harassment, security, and privacy;\\ Ensure the research design is robust and feasible given the online context;\\ Expected outcomes should provide actionable strategies for minimizing harm to the communities/\\online environments, such as debriefing or sandboxing.}\\
     \hline
     Public engagement &
     \shortstack[l]{Consult relevant constituents to identify the emerging issues in the online environment;\\ Ensure results address public concerns and online safety;\\Use feedback from initial public engagement to refine the research approach.}\\
     \hline
     Absence of alternative &
     \shortstack[l]{Demonstrate that no alternative methods can provide equivalent knowledge in terms of validity, scale,\\ or realism;\\ Document how alternatives were considered and why they were judged insufficient.} \\
     \hline
     Appropriate organizer & 
    \shortstack[l]{Researchers should have expertise in large-scale experimental design/human Subjects research;\\ Oversight committees should be in place to monitor the study.} \\
     \hline 
     Informed consent &
     \shortstack[l]{Obtain informed consent wherever possible and follow terms of service, privacy policies, laws; \\ If consent is waived, it should be justified and done so under strict ethical board oversight.}\\
    \hline
     Deception & \shortstack[l]{Use deception only when necessary, and subject it to independent ethical review or oversight, where available;\\Provide a debriefing process after the activity to explain the true nature of the study;\\Allow participants the chance to ask questions during debriefing, either through automated messages \\or direct interaction with a researcher;\\Justify instances where debriefing may not be feasible and seek a waiver if applicable.} \\
     \hline
     Benefits and harms &
     \shortstack[l]{Potential benefits should outweigh risks for society/online communities;\\ Regularly assess harm during the study and establish protocols to mitigate risks;\\Conduct longitudinal studies to evaluate long-term impacts.}\\
     \hline
     Risk minimization &
     \shortstack[l]{Use pilot studies or simulations to identify and mitigate potential risks;\\ Use lowest level of exposure possible that will maintain results;\\ Regularly monitor data security measures to protect against breaches;\\ Avoid targeting specific groups in the online environment.} \\
     \hline
     Third-party risks &
     \shortstack[l]{Ensure adequate safety to avoid any third-party risks in the online environment;\\ Monitor the study's impact to community and mitigate harm as needed;\\Debrief participants.} \\
     \hline
     Selection of participants &
     \shortstack[l]{Ensure diverse representation across demographics and user behaviors;\\Think about what a healthy population means in your context\\Avoid bias in participant recruitment;\\Avoid targeting vulnerable individuals without ethical justification and oversight.} \\
     \hline
     Payment of participants  &
     \shortstack[l]{Ensure payments are sufficient enough to align with risks and outcomes;\\Ensure a balance between anonymity and safety when the study is performed deceptively.} \\ 
     \hline
     Right to withdraw &
     \shortstack[l]{Participants should be able to withdraw at any time without penalty;\\In deceptive online studies, justify the lack of withdrawal opportunities to avoid altering results;\\Offer a debriefing at the end to clarify the study's true purpose.} \\
  \bottomrule
\end{tabular}
\end{table*}

\subsubsection{Informed consent.}

Informed consent is one of the most important ethical issues for conducting research on social challenge studies. The preceding section highlights how online and networked contexts complicate traditional medical models. Informed consent is a process in which potential study participants are provided with clear information about the research study’s purpose, procedures, risks, and potential benefits, after which they may choose to agree to participate or decline \citep{kenneally2012menlo, largent2024ethical}. The purpose of the informed consent procedure is to respect the autonomy of the study participants \citep{kenneally2012menlo}. However, in some cases, when the public interest is at stake, involving study subjects without informed consent may be ethically justifiable. Such an approach is also viable if providing informed consent is not possible and if the participation of the subject does not impact their autonomy. In all cases, the ethical justification of such decisions should be carefully evaluated and reviewed by a research ethics committee, where such review is available \citep{resnik2018ethics,gelinas2016and}.

To guide researchers through the complexity of informed consent, we developed a decision tree (Figure \ref{fig:decisiontree_ethicalFWIC}) in the \textit{Appendix} that outlines the ethical consent pathways for social challenge studies. This flowchart begins by asking whether individual informed consent is possible. If so, researchers are expected to obtain it from participants.  If not, due to study scale, anonymity, or the nature of public platforms, the decision tree outlines structured alternatives. This includes getting consent from a group or community, sending public notices, or seeking a consent waiver through independent ethical review (e.g., IRB/REB), where such mechanisms are available. The tree also guides researchers through cases involving deception or delayed debriefing. It asks them to consider whether participants could be identified, whether they could be harmed, and whether they should be informed about the study later or allowed to opt out.

In these contexts, it is essential to comply with online terms of service, robot.txt files, privacy policies, and applicable regulations and laws. The terms of service and privacy policies represent the baseline agreements that outline how the data of the participants can be used, providing a foundational level of consent for online studies. 

Many social challenge studies, especially those conducted online, often lack informed consent, where study participants are unaware that they are participating in a study. Addressing ethical frameworks for social challenge studies that waive consent due to impracticality is a complex issue that requires careful consideration of participants' autonomy, privacy, and potential harm. Researchers conducting such studies also need to ensure that the data is not identifiable and that the observation of human subjects is indeed in a space where it is permissible to collect data \citep{willis2019observations}.

Another approach is to request broad consent from participants. This means obtaining consent in advance for a variety of research activities without exactly defining the types of data or behaviors that may be examined. For example, participants might agree to contribute to ``research involving social behavior on platforms'' without being informed about the specifics of each individual study. However, it remains an open question as to how to obtain effective consent from groups or participants on a large scale. An option to enhance transparency before conducting the study is to implement widespread public awareness campaigns outlining the purpose, methods, and opt-out options. However, if waiving consent is necessary, the researcher must provide a strong justification by demonstrating that obtaining direct individual consent is not feasible. They must also show that the research offers significant social value, and explain how risks to participants are minimized and justified, typically as part of an independent ethical review where available.

\subsubsection{Deception.}

In some cases, revealing the purpose of a research study can compromise the validity of the results \citep{johnson2011computer}. Consequently, researchers may opt to use deception (if approved by an ethics oversight board) before beginning their study. This deception approval is necessary to protect the interests of the participants and ensure that the research is conducted in an ethical manner. 

Furthermore, most studies that include deception are generally followed by a debriefing process rather than a one-time statement. Once the activity is over, the participants can be informed about the true nature of the study and the reason why deception was necessary. This can be done automatically or with a researcher present to discuss the details and address any questions or concerns. The goal of this process is to ensure that participants leave with a clear picture of their role in the study and a solid understanding of its purpose. However, there has been some criticism of studies that rely on self-discovery as a means of learning about participants' behavior or social interactions \citep{sommers2013forgoing}. In computer security research, if revealing the actual research protocol might cause adverse harm to the participant, and there is minimal risk involved, a waiver for debriefing can be granted by a review board if adequately justified \citep{finn2007designing, johnson2011computer}. 

\subsubsection{Benefits and harms.}  

Clear metrics for assessing the benefits and drawbacks of research are crucial to maintaining ethical integrity, particularly in studies that address social challenges. While social challenge studies offer insights into online behavior, misinformation, and policy effects, the outcomes may not always be as direct or linear. However, they do possess significant potential to improve societal outcomes. To maximize societal benefit, researchers should actively engage with key constituents, including platforms, communities, policymakers, and other relevant parties. By establishing these connections, researchers can ensure that their findings are communicated effectively, thereby increasing their likelihood of impacting decision-making and driving meaningful change. 

Additionally, by involving communities early in the research process, researchers can better align their objectives with the public's needs and expectations. This collaborative approach makes the findings more applicable and fosters trust and transparency, ensuring that the research positively contributes to the broader social context. Responsible storytelling about research outcomes represents another critical balancing act: results should be presented with humility and exactness, while remaining accessible and comprehensible to diverse audiences beyond academia.

Assessing risks in social challenge studies should also consider the possibility of participant or community burdens, such as privacy disruptions or unintended exposure to sensitive content, even when no direct harm occurs. These burdens may be less visible, but they are still important for ethical review and should be weighed together with the study’s potential social benefits. Researchers must navigate the complexities of understanding how participation in these studies affects individuals, groups, or systems, both in the short term and over time. Longitudinal studies that follow up with participants after the conclusion of a study can offer valuable insights into long-term consequences. These types of retrospectives can help identify any unforeseen harms or benefits, as well as enhance the overall understanding of potential risks and harms associated with these types of studies. Currently, there is a lack of sufficient longitudinal studies in this area to fully understand the ripple effects of these studies. Expanding retrospective analyses and conducting meta-studies across various domains could significantly improve our understanding of these risks and help shape ethical guidelines.

Unlike medical challenge studies, social challenge studies often take place in digital, dynamic, decentralized, and only partially controllable environments. As a result, risk–benefit assessment cannot always rely on precise prediction or tightly bounded exposure. The relevant ethical principle is therefore not a utilitarian calculus but a structured and transparent justificatory process. Researchers, therefore, should justify why an intervention is warranted, how they have bounded the scope of exposure, what uncertainties remain, and how emergent effects will be monitored and addressed over time. Risk–benefit reasoning becomes iterative and adaptive rather than determinative. 

While fully definitive metrics may not be immediately feasible, our aim is to call for the development of transparent, evolvable, and accountable approaches for assessing the benefits and potential harms of social challenge studies. More work is needed to create approaches that support ethical justification and provide a foundation for even more future work that further formalizes these evaluations.

\subsubsection{Risk minimization.}

In social challenge studies, potential risks to participants can be minimized by carefully designing the study and ensuring minimal risks to their privacy and well-being. The sensitive data of the study can be protected from misuse by using appropriate data anonymization \citep{zhou2008brief,majeed2020anonymization} and protecting the confidentiality among the team members \citep{nosek2002research}. To avoid any unintended consequences, it is necessary to monitor the study continuously and adapt cybersecurity protocols. The study should be justified ethically to ensure that the benefits of the study outweigh the potential risks.

\subsubsection{Third-party risks.}

Third-party risks in social challenge studies arise when the challenge spreads beyond the intended participants; an example of this is when a participant in a social challenge study is exposed to misinformation and then spreads this information offline through word of mouth. This possible unintended spread is applicable for both types of study cases, where participant selection is feasible, and where participants are anonymous. In both cases, participants can unknowingly involve third parties by sharing the challenge and impacting others. Sometimes participants can also share sensitive and personally identifiable information about third parties without their direct consent \citep{lovato2022limits}. Considering all these risks, researchers should design the study carefully to avoid implicating unintended third parties as much as possible. Debriefing participants, secondary participants, and third parties is another way to clarify the study, explain its true purpose, and mitigate any negative effects.

\subsubsection{Selection of participants.}

The selection of participants for social challenge studies can be largely guided by principles from medical challenge studies, focusing on fairness and minimizing harm. In medical research, participants are often selected from a pool of healthy young adults, as they are generally less likely to experience severe adverse effects. When applying this approach to online environments, a key question arises: What defines a ``healthy'' participant or group in a virtual context? Identifying criteria for individuals or groups capable of enduring challenging conditions without harm is a critical area for further research and an important question for online social networks. Developing robust measures to assess this resilience is crucial to ensure the ethical selection of participants in these studies.

Researchers may also want to consider the diversity of online environments and the unique vulnerabilities that participants may face. This includes evaluating whether groups are adequately prepared to handle the risks associated with the study while minimizing the chances of harm. Establishing clearer guidelines and metrics for participant selection in social challenge studies is important to maintain ethical standards and protect the welfare of both individuals and groups.

Although participant selection guidelines in medical challenge studies often reference the exclusion of vulnerable populations, the concept of vulnerability itself is the subject of substantial debate in bioethics and the social sciences. Scholars have shown that vulnerability is not a fixed property of particular groups but a contextual and relational condition shaped by structural inequities, institutional dynamics, and features of the research environment \cite{van2018vulnerability, rogers2014vulnerability}. Recognizing this complexity is essential for both medical and social challenge studies. Rather than presuming that certain populations are intrinsically vulnerable, researchers should specify the contextual factors that might limit autonomy, increase exposure to harm, or undermine voluntariness in their particular study. This approach shifts the focus from categorizing groups to identifying and mitigating the situational sources of vulnerability that can arise in challenge study designs.

\subsubsection{Payment of participants.}

In medical challenge studies, payment to participants is a means of compensating them for their time and contributions to the study. Compensation procedures can vary according to the type of study. For social challenge studies, such as surveys or online research where participant selection is possible, offering payment can effectively compensate individuals and recognize their contributions to the research. This payment should be sufficient to align it with the risks and outcomes. The payment amount should not be too high so that the vulnerable group will be influenced by it and participate in the study without considering any risks and consequences. However, compensating participants may be impractical in online studies where it is not possible to select specific individuals. Participants in such studies are mainly anonymous and unidentifiable, which complicates the process of offering them payment. Unlike medical challenge studies, which typically have straightforward methods for compensating participants, online research must find a balance between maintaining participants' anonymity and ensuring their safety.  

\subsubsection{Right to withdraw.}

The right to withdraw is an important ethical aspect of medical challenge studies, where participants have the freedom to leave the study at any time without negative consequences. However, in anonymous online social challenge studies, this right may be difficult to achieve. In some cases, participants may not know about their participation in studies, which makes it challenging to exercise their right to withdraw. In such cases, researchers must provide a clear ethical justification for using deception and waiving consent in the study and explain why withdrawal from the study is not possible. They must ensure that this practice is crucial to achieving research goals while preventing harm to individuals. Ethical oversight boards can play a crucial role in assessing the ethical justification by ensuring a balance between potential benefits and the ethical concerns associated with limiting participant autonomy. Researchers should consider additional ethical safeguards, such as data security and providing a clear debriefing after the study to protect participants' rights and well-being.

\section{Conclusion}
\label{sec:discussion}
This work aims to define social challenge studies in order to highlight lessons and emerging ethical challenges based on existing frameworks from clinical research. Drawing from case studies, we proposed general methodological guidelines that prioritize participant autonomy, informed consent, and societal benefits. We support the use of longitudinal studies to investigate the long-term effects of participation in social challenge studies that have not been thoroughly examined. Retrospective analyses and meta-studies across fields could also uncover unforeseen harms and refine ethical guidelines.

Several open questions remain. One of the primary challenges is to effectively obtain large-scale consent in online and virtual environments, where the indirect and diffuse nature of potential harms complicates traditional models of individual consent and risk assessment. Additionally, clarifying what it means for a participant or community to be ``healthy'' or ``resilient'' in virtual contexts is essential and warrants careful study to ensure appropriate participant selection and quantify potential harms. Part of the issue lies in identifying measures of health, resilience, and risks, which are domain-specific but should nevertheless be the subject of future research.
In addition, future domain-specific studies should also develop a more structured and systematic approach for understanding the appropriate dose-response relationship that balances risk in social challenge studies.

We view this paper as the beginning of an important conversation. In future work, we will refine these guidelines in collaboration with a multidisciplinary group of researchers to develop clearer measures for the aforementioned open questions. We also plan to engage in longitudinal analysis on the benefits and harms of social challenge studies. Our ultimate goal is to encourage discussion about ethical guidelines for safely expanding the applicability of social challenge studies, ensuring the well-being of their participants while promoting research for social good.

\section{Educational Implications}
The growing body of work and the ethical issues surrounding social challenge studies underscore the need to educate both current and future researchers, as well as the ethical review boards, on ethical best practices. Training programs in computational social science, computer science, and data science rarely include structured modules on research ethics beyond standard human subjects protections. However, as demonstrated in case studies of misinformation exposure, bot deployment, and adversarial interventions, researchers routinely face decisions analogous to those in medical challenge trials, which are not typically covered in standard ethics training or in course curriculum. College curricula could also work to incorporate case-based instruction on social challenge studies, paralleling medical ethics education in human challenge research. Such modules would foster researcher competence in risk-benefit assessment, consent in digital environments, and the management of community-level harms. Beyond formal curricula and professional development workshops, continuing education for researchers working in digital and sociotechnical fields should provide targeted training in ethical review, consent practices, and post-study debriefing processes.

\section{Best Practices}
We are drawing on best practices from the medical field to suggest ethical frameworks for computational social science researchers who conduct social challenge studies. The best practices paralleled from medical challenge studies can be summarized as follows. First, transparency about the risks and benefits in the study is necessary. Researchers should clearly state in their information sheets and consent protocols why exposing participants or communities to potential harm is justified and demonstrate the absence of reasonable alternatives. Second, consent and debriefing need clear processes and approvals, especially when seeking exemptions. Individual or group consent in online contexts should be explicit, and mandatory debriefing should be provided whenever deception or covert exposure is used. Third, independent oversight should be incorporated into the process. Institutional review boards and research ethics committees could establish specialized review processes for social challenge studies, paralleling similar committees in medical research. 

In addition, risk minimization should guide research design and favor longitudinal studies that monitor effects on participants as well as methods that reduce unnecessary exposure to harm. Equity in participation is also critical: researchers should avoid disproportionately targeting vulnerable or high-risk populations. Ultimately, transparency and public engagement should also be established as routine practices. Engaging communities that may be affected by interventions and openly sharing findings can help maintain trust and accountability.

\section{Research Agenda}
This study presents multiple open questions and directions for future empirical and theoretical work on the ethics of social challenge research. Further work needed in this area is as follows: 1.) Measure long-term harms through longitudinal studies that assess whether brief exposure to misinformation, bots, or adversarial interventions leaves lasting effects on individuals or communities. 2.) Address third-party or secondary participant risks, particularly how online communities experience unintended consequences of research interventions. 3.) Develop and test consent models, especially in networked or grouped environments. 4.) Investigate what it means for a participant or community to be ``healthy'' in virtual contexts. 5.) Establish a structured approach for understanding what is a proper dosage and the dose-response relationship that balances risk in social challenge studies. 6.) Study how research ethics committees, journals, news, public opinion, and funding agencies respond to social challenge research, which could provide valuable insight into governance gaps. 7.) Pursue empirical work on educational outcomes, assessing whether training on the ethics of social challenge studies influences researchers' decision-making. These open questions are important to build a robust ethical framework for social challenge studies and will help ensure both the protection of participants and the advancement of socially valuable research in this area.

\bibliographystyle{SageH} 
\bibliography{bib.bib}

\appendix

\section{Appendix I: Decision Trees}
\label{sec:appendixi}
\begin{figure*}[t]
\centering
\includegraphics[width=0.9\textwidth]{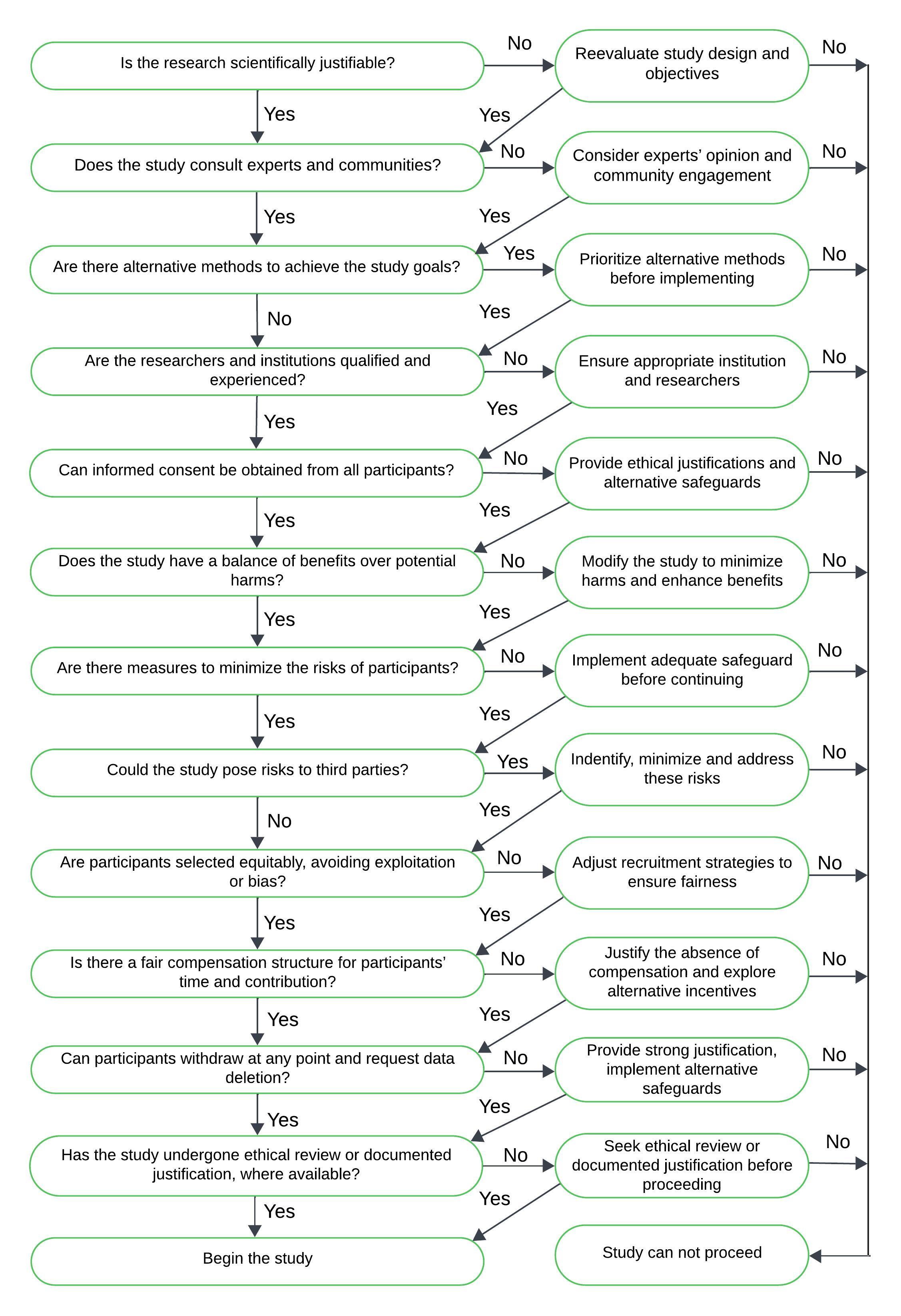}
\captionsetup{format=plain, justification=justified, singlelinecheck=false}
\caption{Decision tree outlining steps in the social challenge study ethical framework. This decision tree supports researchers in systematically evaluating key ethical considerations when designing or reviewing social challenge studies. It outlines steps, including assessment of exposure risks, consent pathways, participant vulnerability, use of deception, and harm mitigation strategies. Drawing analytically on medical challenge frameworks, the tree is designed to guide ethical decision-making for studies involving bots, misinformation, platform interventions, honeypots, and other controlled-risk designs in digital environments.}
\label{fig:decisiontree_ethicalFW}
\end{figure*}

\begin{figure*}[t]
\centering
\includegraphics[width=0.9\textwidth]{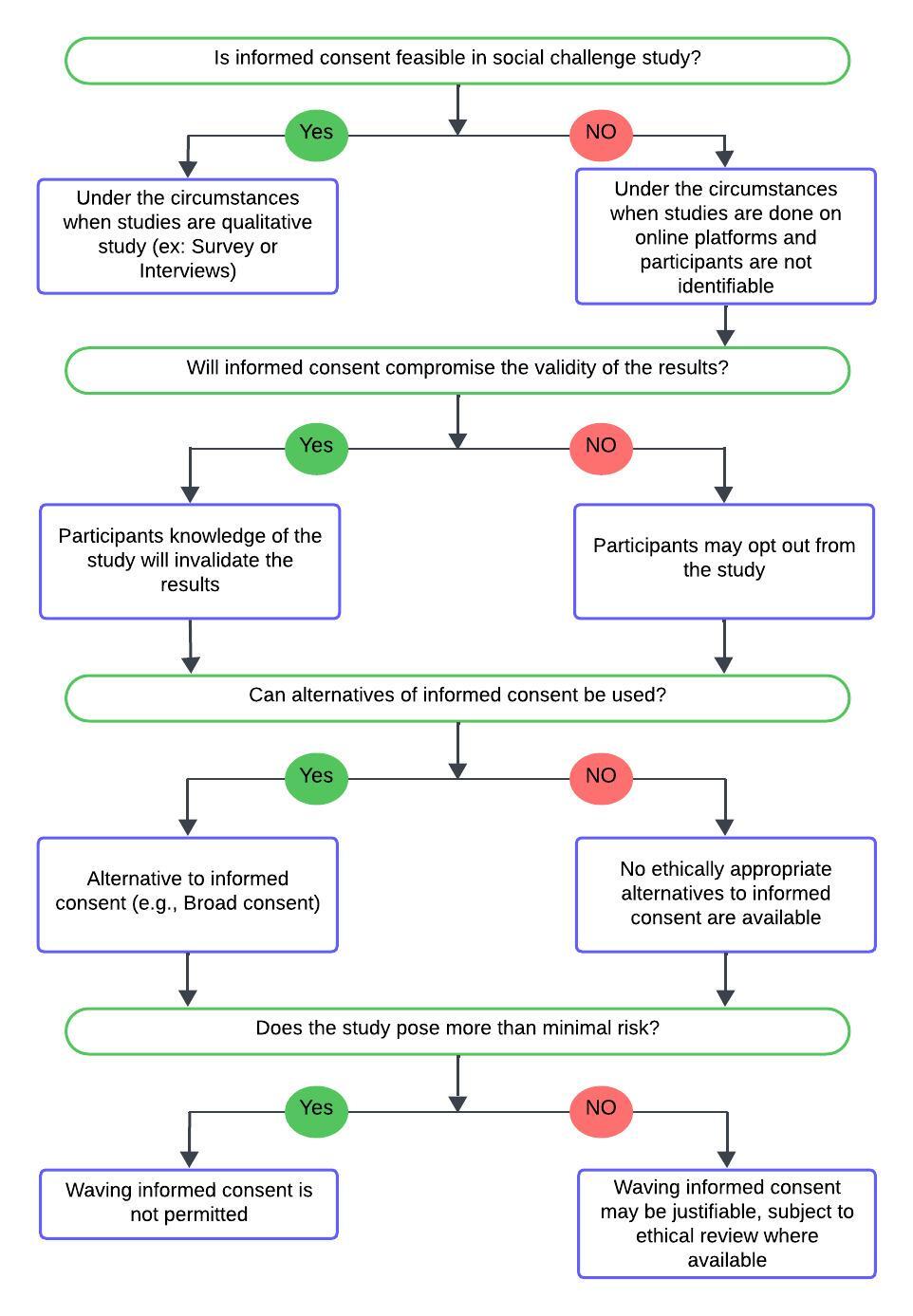}
\captionsetup{format=plain, justification=justified, singlelinecheck=false}
\caption{Decision tree outlining steps in the social challenge study ethical framework related to informed consent. This tree helps researchers assess how informed consent should be obtained, waived, or substituted in studies involving deception, public interactions, or group-level exposure. It highlights the role of terms of service, the feasibility of direct consent, the importance of debriefing when possible, and ethical oversight for waivers or documented ethical justification for waivers, where applicable. It supports researchers conducting studies in environments where traditional consent is difficult or impossible to implement.}
\label{fig:decisiontree_ethicalFWIC}
\end{figure*}

The following decision trees (Figs.~\ref{fig:decisiontree_ethicalFW} and \ref{fig:decisiontree_ethicalFWIC}) are intended to help researchers, and where applicable, ethics committees or IRB in applying our proposed ethical framework to social challenge studies. These tools support structured decision-making related to exposure risk, consent, deception, harm mitigation, and oversight. We recognize that not all researchers have access to formal ethics review bodies; in such cases, ethical responsibility remains with researchers to adhere to shared ethical frameworks and professional norms.

\end{document}